\documentclass[pre,preprint]{revtex4}
\usepackage[dvipdfm]{graphics}
\usepackage{subfigure}
\begin{document}
\title{Frequency synchronization in random oscillator network}
\author{Takashi Ichinomiya}
\affiliation{Meme Media laboratory, Hokkaido Univ. Sapporo, Hokkaido, Japan}
\email{miya@aurora.es.hokudai.ac.jp}
\date{12 Nov 2003}
\begin{abstract}
 We study the frequency-synchronization of  randomly coupled oscillators. By
 analyzing the continuum limit, we obtain the sufficient condition for the
 mean-field type synchronization.  We especially find  that  the critical
 coupling constant $K$ becomes 0 in the random scale free network,
 $P(k)\propto k^{-\gamma}$, if $2 < \gamma \le 3$. Numerical simulations
 in finite networks are consistent with these analysis. 
\end{abstract}
\maketitle
\section{Introduction}
 Recently, it has become clear that the complex network plays an important
 role  in  many natural and artificial systems, such as neural network,
 metabolic systems, power supply system, Internet, and so on
\cite{Albert,Dorogovtsev}.
 In particular, we have recognized that many networks have scale-free topology;
 the distribution of the degree  obeys the power law,
 $P(k)\sim k^{-\gamma}$. 
 The study of the scale-free network now attracts the interests of many
 researchers in mathematics, physics, engineering and biology. 

 The dynamics in the network systems  is  one of the important themes of
 the investigation of complex network. 
 In this paper, we study the synchronization of the random network of
 oscillators.  The phase synchronization in complex network has been
 studied by several authors\cite{Timme, Wang}, while 
 the frequency synchronization has not been studied so much.
 One of the important work on this problem  was made by
 Watts\cite{Watts}. He suggested from numerical simulation that
 mean-field type synchronization
 occurs in small-world network such as Watts-Strogatz model. His study was
 followed  by the work of Hong {\it et al.}, in which phase-diagram and
 critical exponent are numerically studied in detail\cite{Hong}.
 These works showed that mean-field type synchronization, that
  Kuramoto observed in globally-coupled oscillators\cite{Kuramoto}, appears
 also in the  small-world network. However, such a study in scale-free
 network has not been performed yet.

 In this paper,  we analytically study the frequency synchronization 
 in the random network of oscillators.  By analyzing the
  continuum limit of this model, we obtain  the sufficient condition for
 the synchronization.
  Our result shows that  in the scale-free random network the threshold for
 synchronization is absent if $2 < \gamma \le 3$. We also carry out the
 numerical simulations and the results are consistent with this analysis.

This paper is constructed as follows: the next section describes the
model of oscillator network  and derive  the continuum
limit equation. Section \ref{analysis} is devoted to derive the
sufficient condition for the synchronization from the continuum limit 
equation. We show that the order parameter is different from the one 
used in previous works, and especially we conclude that  the threshold
for the synchronization disappears in the random
scale-free network. These
results are consistent with the results of the numerical simulations,
which are described in 
section \ref{simulation}. In the last section we make a summary of this
paper and discuss the relation to the other properties of the scale-free
 network.

\section{Oscillator network model and its continuum limit}

  First we describe the model we study in this paper. We study the network
 with N-nodes. At each nodes, there exists a oscillator and  the phase of the
 oscillator $\theta_i$ is developed as
\begin{equation}
 \frac{\partial \theta_i}{\partial t}=\omega_i + K\sum_j a_{i,j}\sin(\theta_j-\theta_i).\label{problem}
\end{equation}
  where $K$ is the coupling constant,  $a_{i,j}$ is 1 if the nodes
 $i$ and $j$ are connected, and 0 otherwise. $\omega_i$ is a random
 number, whose distribution is given by the function $N(\omega)$.
 
 For the analytic study, it is convenient to use the continuum limit 
 equation.  We define $P(k)$  as the
 distribution of nodes with degree $k$,  and  $\rho(k,\omega; t,\theta)$
 the  density of oscillators with phase $\theta$ at time $t$, for given
 $\omega$ and $k$.
We assume that $\rho(k,\omega; t,\theta)$ is normalized as 
\begin{equation}
 \int_0^{2 \pi} \rho(k,\omega; t, \theta) d \theta = 1.
\end{equation}
 For simplicity, we assume $N(\omega)=N(-\omega)$. Under this assumption,
 we suppose that the collective oscillation  corresponds to the 
 stable solution,
 $\frac{d \rho}{dt} = 0$, in this model.

 Now we  construct the continuum limit equation for the  network of
 oscillators.
 The evolution of $\rho$ is determined by the continuity equation $\partial \rho /\partial t = -\partial (\rho v)/\partial \theta$, where $v$ is defined by
 the continuum limit of the r.h.s of eq.(\ref{problem}). Because one randomly 
selected edge connects to the
 node of  degree $k$, frequency $\omega$, phase $\theta$ with the
 probability $kP(k)N(\omega)\rho(k,\omega;t,\theta)/\int dk k P(k)$,
 $\rho(k,\omega; t, \theta)$
 obeys the equation
\begin{equation}
  \frac{\partial \rho(k,\omega; t,\theta)}{\partial  t} = -\frac{\partial}
{\partial \theta}\left[\rho(k,\omega;t,\theta)\left(\omega+ \frac{Kk\int
 d\omega^{\prime}\int dk^{\prime}\int d \theta^{\prime}N(\omega^{\prime})
P(k^{\prime})k^{\prime}
 \rho(k^\prime,\omega^{\prime};t,\theta^{\prime})\sin(\theta-\theta^{\prime})}{\int dk^{\prime} P(k^{\prime})k^{\prime}}\right)\right].\label{master}
\end{equation}
In the next section, we study the mean-field solution of this equation.

\section{Mean-field analysis of random oscillator network\label{analysis}}

In this section, we  study the sufficient condition for the
synchronization  using   eq.(\ref{master}).
  First we introduce order parameter $(r,\psi)$ as 
\begin{equation}
  re^{i \psi} = \int d\omega \int dk \int d\theta N(\omega)P(k)k \rho(k,\omega;t,\theta) e^{i \theta}/\int dk P(k)k.\label{mf}
\end{equation}
 This order parameter is different from the one which is used in previous work 
in small-world model\cite{Watts,Hong}. In the previous works,
 $\sum_i e^{i \theta_i}/N$ is used for the mean-field, while 
 our order parameter corresponds to $\sum_i k_i e^{i \theta_i}/\sum_i k_i$,
 where $k_i$ is the degree of the node $i$.
 However, from eq.(\ref{master}) it seems natural to use eq.(\ref{mf}) as the
 mean-field value the in random network.  Here we note that $0\leq r \leq 1$.

Inserting eq.(\ref{mf}) into eq.(\ref{master}), we get
\begin{equation}
   \frac{\partial\rho(k,\omega; t,\theta)}{\partial t} =-\frac{\partial}{\partial \theta}(\rho(k,\omega;t,\theta)(\omega + K k r \sin(\psi-\theta))).\label{mf-eq}
\end{equation}
 The time-independent solution of $\rho$ is then
\begin{equation}
  \frac{\partial}{\partial \theta}(\rho(k,\omega;t,\theta)(\omega + K k r \sin(\psi-\theta))) = 0.
\end{equation}
 Without a loss of generality,  we can  assume  $\psi=0$. 
Since we want to seek the solution which corresponds to the Kuramoto's solution
 in globally coupled oscillators, we assume the solution of this equation as
\begin{equation}
 \rho(k,\omega;\theta)=\left\{
                        \begin{array}{ll}
                         \delta(\theta - \arcsin(\frac{\omega}{Kkr})) & \mbox{ if } \frac{|\omega|}{Kkr} \le 1 \\
                         \frac{C(k,\omega)}{|\omega-Kkr\sin\theta|} & \mbox{ otherwise,}
                        \end{array}
                       \right.\label{density}
\end{equation}
where $C(k,\omega)$ is the normalization factor. Here we note that $\rho$
depends on both  $K$ and  $k$. This equation means that $Kk$
corresponds to the coupling between mean-field and the oscillator. 
Inserting eq.(\ref{density}) into eq.(\ref{mf}), we get the equation for $r$,
\begin{equation}
 r = \int d\omega\int dk\int d\theta N(\omega)kP(k)\rho(k,\omega;\theta)e^{i \theta}/\int dk k P(k).
\end{equation}
 To calculate this integral, first we divide the integral over $\omega$.
 \begin{eqnarray}
  \int d\omega\int dk\int d\theta
   N(\omega)kP(k)\rho(k,\omega;\theta)e^{i \theta} &=& \int dk \int d
   \theta
   \left(\int_{-Kkr}^{Kkr}d\omega+\int_{-\infty}^{-Kkr}d\omega+\int_{Kkr}^{\infty}
    d\omega\right)\nonumber\\
&\times& N(\omega)kP(k)\rho(k,\omega;\theta)e^{i\theta} \label{mf-eq-2}
 \end{eqnarray}
  The contribution from  the integral at $\omega< -Kkr$ and
 $\omega > Kkr$ is 0 if $N(\omega)= N(-\omega)$, because 
\begin{equation}
 (\int_{-\infty}^{-Kkr} d \omega+\int_{Kkr}^{\infty}d \omega )
 N(\omega)\rho(k,\omega;\theta)e^{i\theta}=\int_{Kkr}^{\infty}N(\omega)e^{i\theta}C(k,\omega)(\frac{1}{\omega-Kkr\sin\theta}+\frac{1}{\omega+Kkr\sin\theta}).\label{for-0}
\end{equation}
 The integral  of the r.h.s of eq.(\ref{for-0}) over $\theta$ equals to 0.
 Therefore eq.(\ref{mf-eq-2}) is equivalent to
\begin{equation}
 r = \int dk\int_{-Kkr}^{Kkr} N(\omega)k P(k)\exp(i \arcsin(\frac{\omega}{Kkr}))/\int dk k P(k).
\end{equation} 
If we assume $\arcsin(\frac{\omega}{Kkr})$ is between $[-\pi/2,\pi/2]$, we get
\begin{eqnarray}
 r \int dk k P(k) &=& \int dk \int_{-Kkr}^{Kkr} d\omega N(\omega)k P(k) \sqrt{1-\left(\frac{\omega}{Kkr}\right)^2}\nonumber\\
&=& \int dk \int_{-1}^{1} d\omega^{\prime}k P(k)N(Kkr\omega^{\prime})\sqrt{1-\omega^{\prime 2}}\times Kkr\nonumber\\
&=& Kr \int dk k^2 P(k)\int _{-1}^{1}d\omega^{\prime}N(Kkr\omega^{\prime})\sqrt{1-\omega^{\prime 2}}.\label{finaleq}
\end{eqnarray}
If $r \neq 0$, we get 
\begin{equation}
\int dk k P(k) = K \int dk k^2 P(k)\int _{-1}^{1}d\omega^{\prime}N(Kkr\omega^{\prime})\sqrt{1-\omega^{\prime 2}}
\label{final2}
\end{equation}
  The l.h.s of this equation is independent of $r$ and we define the
 r.h.s of this equation as $f(r)$.  
 At $r=1$, $f(r)$ is not larger than $\int d k k P(k) $, 
 because
\begin{eqnarray}
 \int dk k^2 P(k)\int_{-1}^{1} d\omega^{\prime} N(Kkr)\sqrt{1-\omega^{\prime 2}}
 &\le& \int dk k^2 P(k)\int_{-1}^{1} d\omega^{\prime} N(Kkr\omega^{\prime})\nonumber\\
 &\le& \int dk k^2 P(k) \frac{1}{Kkr}\int_{-\infty}^{\infty} d\omega^{\prime\prime}N(\omega^{\prime\prime})\nonumber\\
 &=& \frac{\int dk kP(k)}{Kr}, 
\end{eqnarray}
 here we use the relation $\int_{-\infty}^{\infty} d\omega N(\omega)=1$.
 Therefore
 the sufficient condition that eq.(\ref{finaleq}) have solution at 
 $0 < r \le 1 $
 is that  $f(r) > \int dk k P(k)$ at $r=0$,
\begin{equation}
\frac{KN(0)\pi\int dk k^2 P(k)}{2 \int dk k P(k)} > 1.\label{critical}
\end{equation}
This is the sufficient condition for the synchronization in random
 network of oscillators. 
 The most impressive point of this equation is that in the random scale-free
 network, $P(k)\propto k^{-\gamma}$,
 this condition is satisfied  for any $K > 0$ if $2 < \gamma\le 3$, because
 $\int dk k^2 P(k)/\int dk k P(k)$ diverges. Therefore  we have no
 threshold for the synchronization in the random scale-free
 network. This seems similar to the absence of the threshold in
 susceptible-infected-susceptible(SIS) model\cite{SIS0}. We will discuss
 on this similarity in later section.

In this section, we derive the sufficient condition for the
synchronization in random network of oscillators, using the continuum
limit equation. In the next section, we show that the analysis above is
in good agreement with the results of the numerical simulations.

\section{Numerical simulation of synchronization\label{simulation}}
 In this section, we show the result of the numerical simulations of the
 random network of oscillators. In the all simulations, 
 we take $N(\omega)$ as  $N(\omega)=0.5$ if  $-1.0< \omega< 1.0$, and 0
 otherwise.

 First we show the result on the 1000-node
 Erd\H{o}s-R\'enyi random network model. 
 We choose the probability of coupling $p=0.005$, which gives 
 $\int dk kP(k)=5.0$ and  $\int dk k^2 P(k) = 29.7$
 on average. 
 In this case,  estimated critical $K$ is $K_c =0.214$.  Each
 simulation is carried out 100 times.

\begin{figure*}[t]
\mbox{
% \subfigure[$K=0.15$]{\resizebox{.4\textwidth}{!}{\includegraphics{k0.15wv-random.eps}}}
% \subfigure[$K=0.30$]{\resizebox{.4\textwidth}{!}{\includegraphics{k0.30wv-random.eps}}
 \subfigure[$K=0.15$]{\resizebox{.4\textwidth}{!}{\includegraphics{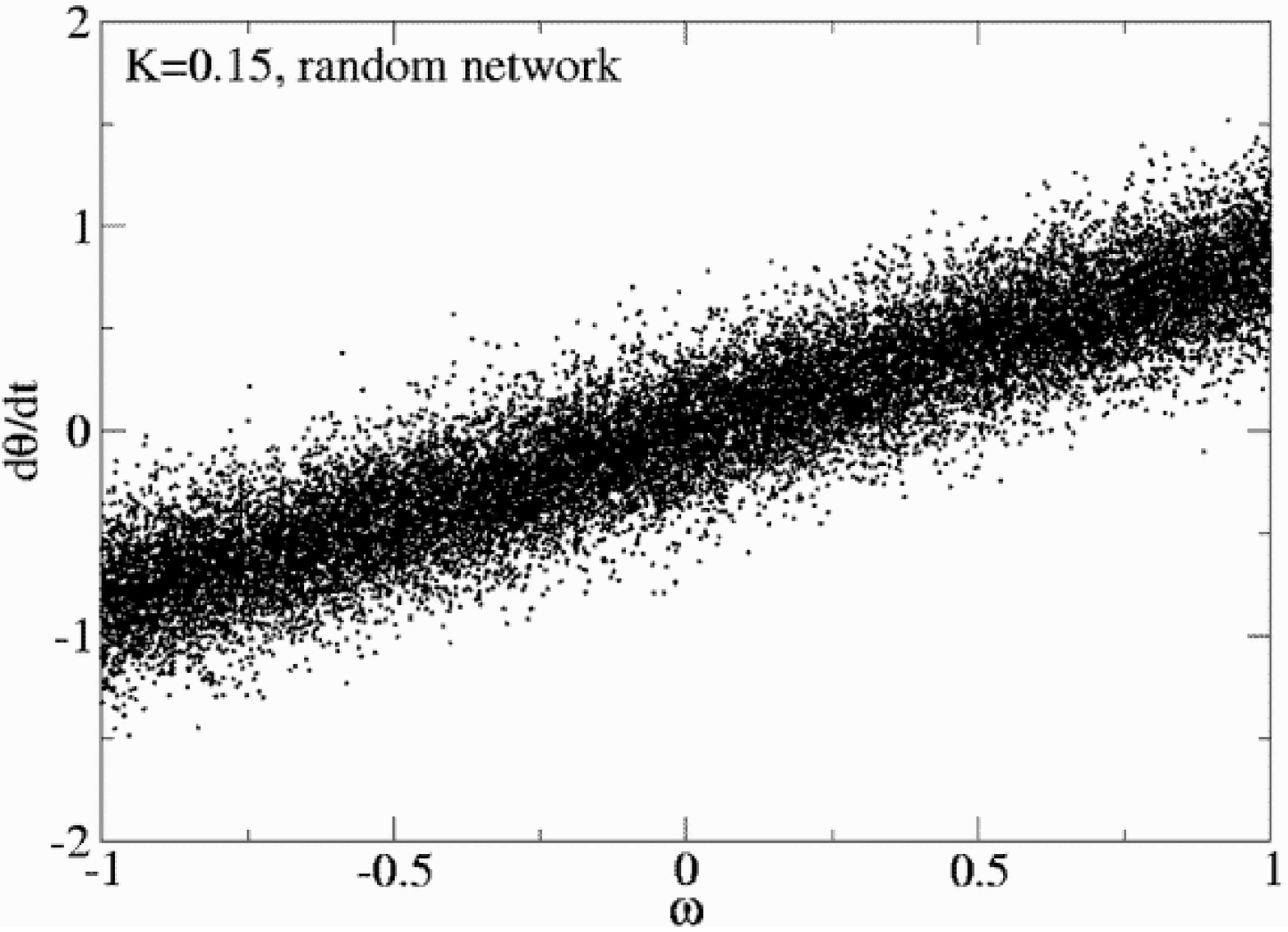}}}
 \subfigure[$K=0.30$]{\resizebox{.4\textwidth}{!}{\includegraphics{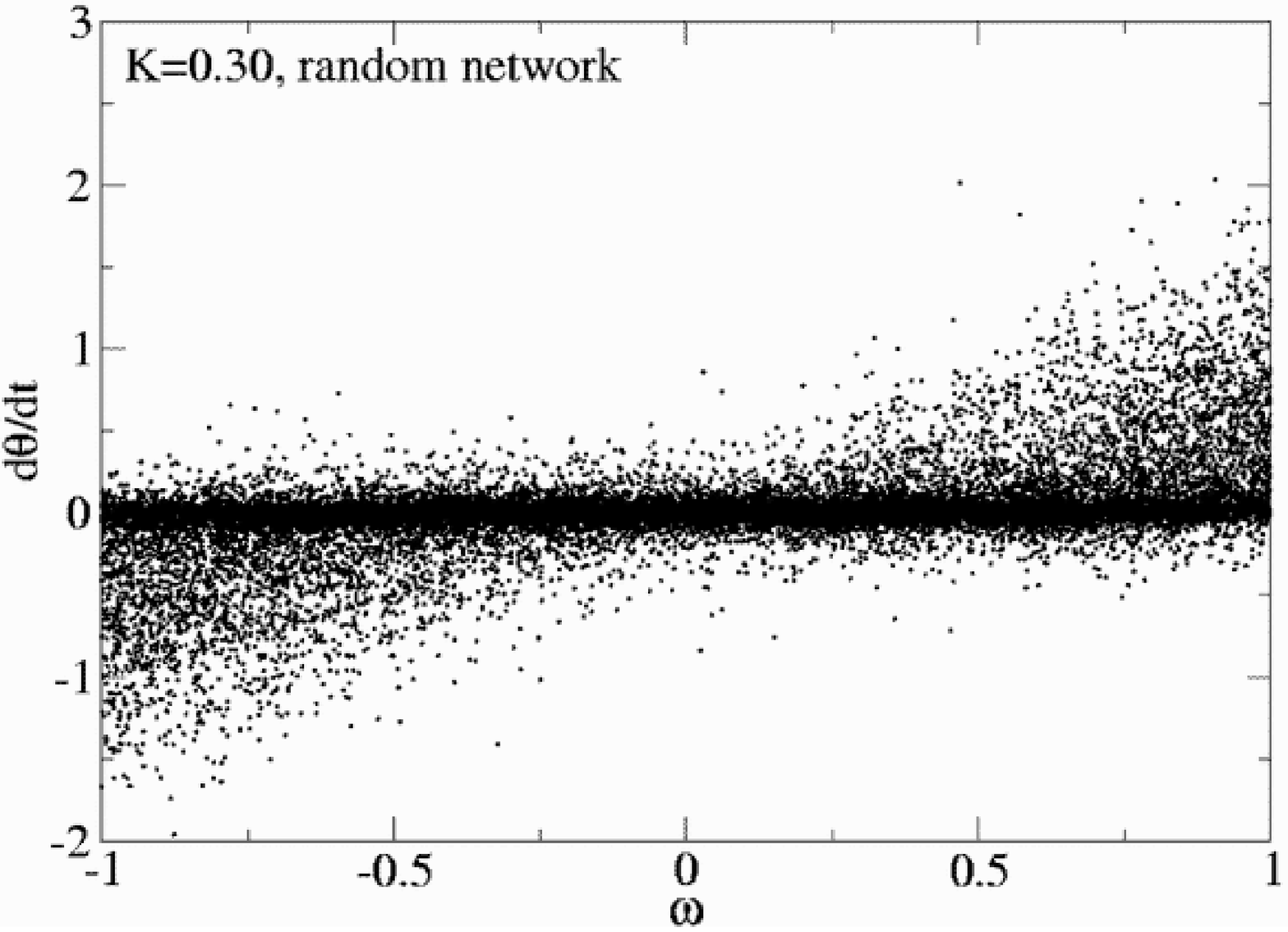}}
}
}

  \caption{$(\omega,d\theta/dt)$ distribution of oscillators  in random
 network, for K=0.15 and K= 0.30.\label{omega-vel-randomnet}}
\end{figure*}

 In Fig.\ref{omega-vel-randomnet} we plot the relation between $\omega_i$ and
 $\frac{d \theta_i}{dt}$ after a long time ($t=200$) when $K=0.15$ and $0.30$.
 In the case of $K=0.15$, $d\theta/dt$ seems to depend on $\omega$ linearly.
 On the other hand at $K=0.30$ many oscillators seem to be synchronized at 
$d\theta/dt = 0$. This figure strongly suggests that the synchronization occurs
 between $K=0.15$ and 0.30.

\begin{figure*}[t]
\mbox{
%   \subfigure[$K=0.15$]{\resizebox{.4\textwidth}{!}{\includegraphics{k0.15-random.eps}}} 
% \subfigure[$K=0.30$]{\resizebox{.4\textwidth}{!}{\includegraphics{k0.30-random.eps}}}
   \subfigure[$K=0.15$]{\resizebox{.4\textwidth}{!}{\includegraphics{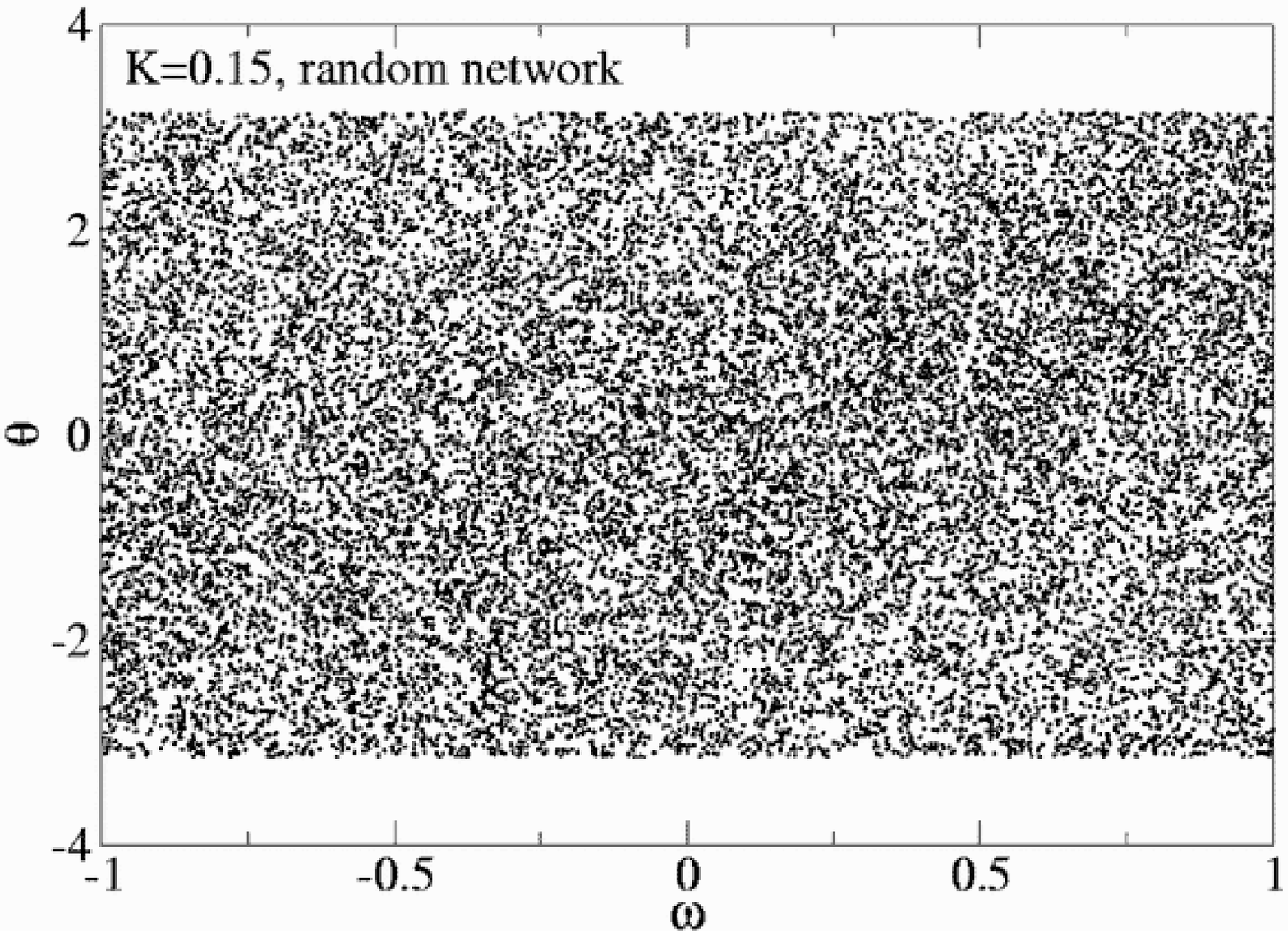}}} 
 \subfigure[$K=0.30$]{\resizebox{.4\textwidth}{!}{\includegraphics{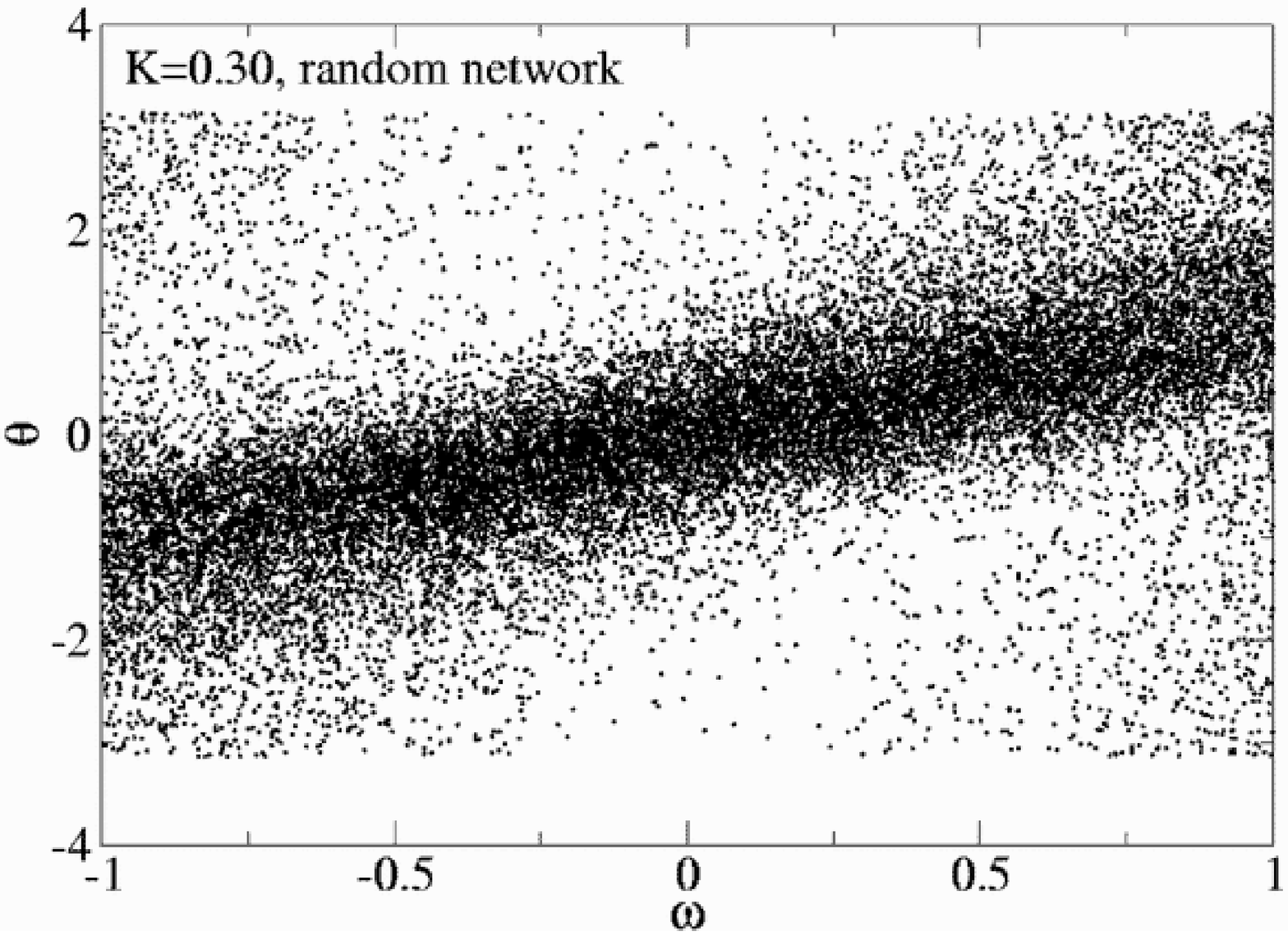}} }
}
  \caption{$(\omega,\theta)$ distribution of oscillators  in random network, K=0.15 and K= 0.30.\label{omega-theta-randomnet}}
\end{figure*}

\begin{figure*}[t]
\mbox{
% \subfigure[$k=3$]{\resizebox{.3\textwidth}{!}{\includegraphics{k0.30-random-k3.eps}}}
% \subfigure[$k=5$]{\resizebox{.3\textwidth}{!}{\includegraphics{k0.30-random-k5.eps}}}
% \subfigure[$k=7$]{\resizebox{.3\textwidth}{!}{\includegraphics{k0.30-random-k7.eps}}}
 \subfigure[$k=3$]{\resizebox{.3\textwidth}{!}{\includegraphics{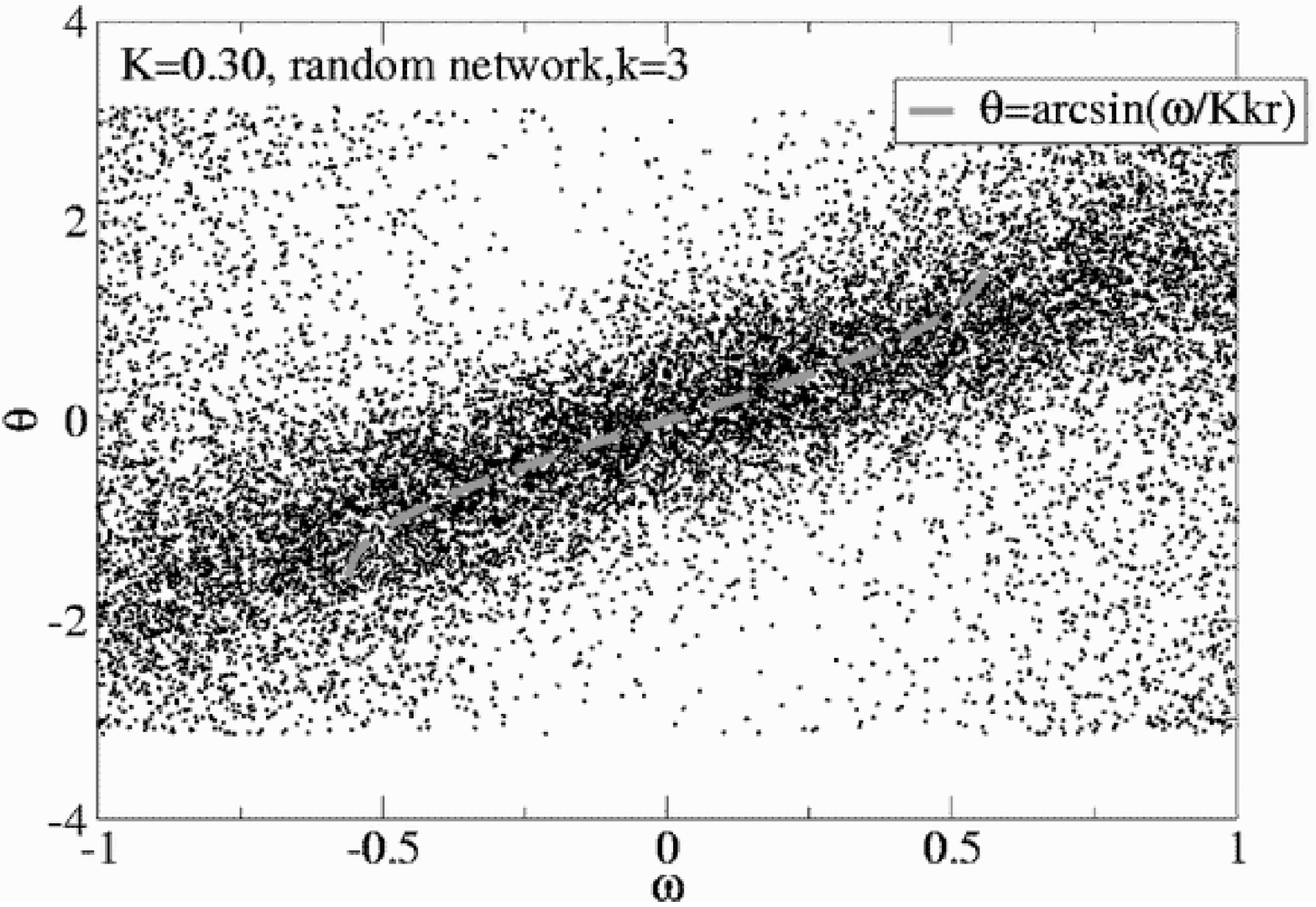}}}
 \subfigure[$k=5$]{\resizebox{.3\textwidth}{!}{\includegraphics{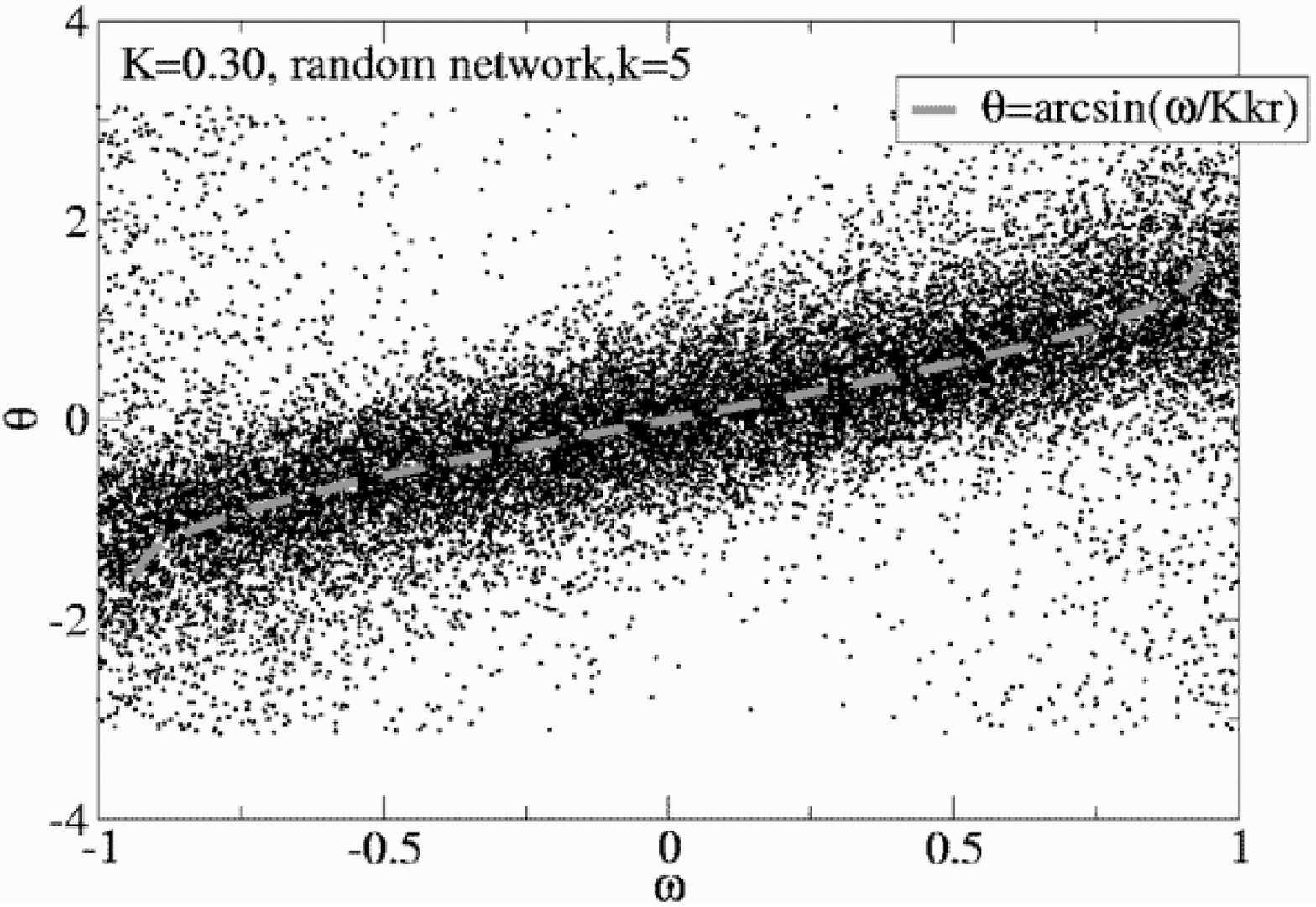}}}
 \subfigure[$k=7$]{\resizebox{.3\textwidth}{!}{\includegraphics{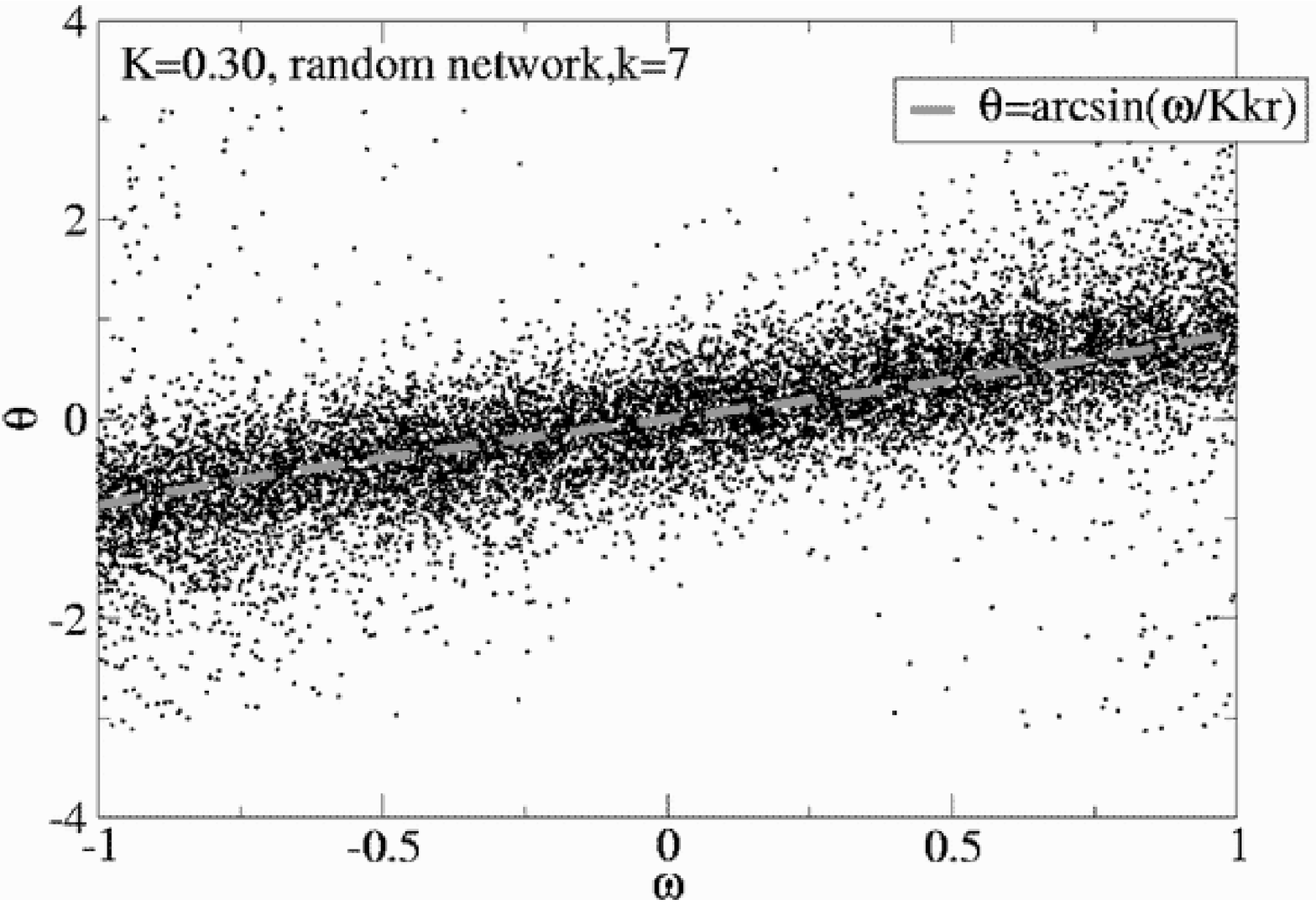}}}
}
 \caption{$(\omega,\theta)$ distribution of oscillators with degree 3, 5 and 7 
  in random network, K= 0.30.\label{omega-theta-randomnet-k}}
\end{figure*}

  We plot the relation between $\omega$
 and $\theta$ for $K=0.15$ and $K=0.30$ in
 Fig. \ref{omega-theta-randomnet}.  We find a clear difference
 between these two cases. In the case of $K=0.30$, the distribution of
 $(\omega_i,\theta_i)$ is apparently non-uniform, while at $K=0.15$ we
 cannot find any structure. In the case of $K=0.30$, $\theta$ seems to depend
 linearly on $\omega$. However, from the previous analysis we suggest that
  $\theta$ depends on both $\omega$ and $k$. To clarify
 the degree dependence,  we plot $(\omega,\theta)$ for the nodes with
 the degree $k=$ 3, 5 and 7  at $K=0.30$ in Fig. \ref{omega-theta-randomnet-k}
. We also plot $\arcsin(\omega/Kkr)$ in these figures.  The average  of
 $r$ is
 0.623 in our simulation. From these figures, we find that
 distribution of $(\omega,\theta)$ seems to be  concentrated around a  single
 line. The concentration line qualitatively coincides with
 $\theta=\arcsin(\omega/Kkr)$. This result suggests that  our mean-field
 defined by eq.(\ref{mf}) is the correct one.

\begin{figure}[t]
%  \resizebox{.4\textwidth}{!}{\includegraphics{mfvalue.eps}}
  \resizebox{.4\textwidth}{!}{\includegraphics{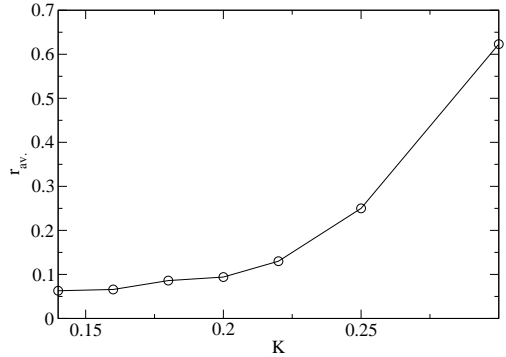}}
  \caption{Interaction dependence of mean-field parameter $r_{\mbox{av.}}$.\label{K-r}}
 \end{figure}

 To estimate the critical coupling $K_c$, we plot $K$ dependence of the 
average of the order parameter $r_{\mbox{av.}}$ in
 Fig.\ref{K-r}. $r_{\mbox{av.}}$ is less than 0.1
 and shows weak dependence on $K$ at $K<0.2$. This 
 non-zero value of $r_{\mbox{av.}}$ is due to the finite-size effect.
 On the other hand, at $K>0.2$, $r_{\mbox{av.}}$
 increases rapidly as  the interaction increases. This figure suggest that 
 $K_c$ is about 0.2, which is in agreement with our
 analysis. Therefore we conclude that all numerical results are consistent with
 our analysis.

\begin{figure}[t]
  \resizebox{.4\textwidth}{!}{\includegraphics{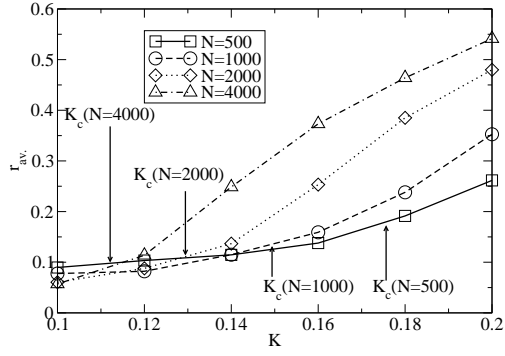}}
  \caption{Interaction dependence of the mean-field parameter in the random
  scale-free network for N=500,1000,2000 and 4000. The arrow shows $K_c$ estimated from
  eq.(\ref{finaleq}). Simulations for each parameters are carried out at
  least 50 realization of the networks.\label{scalefree-r}}
 \end{figure}

 From these simulation, we find that our mean-field theory is applicable to
 Erd\H{o}s-R\'{e}nyi model. However, the most impressive suggestion of our
 analysis is the absence of the threshold in the random scale-free
 network. In the following, we show the result of simulation in random
 scale-free network with $\gamma=2.5$. 
 In Fig. {\ref{scalefree-r}}, we show the relation between order
 parameter $r$ and coupling constant $K$ for
 N=500, 1000, 2000 and 4000. At  N=500 , $r_{\mbox{av.}}$ rapidly increase above
 $K\sim 0.16$, which is 
 qualitatively consistent with the $K_c\sim 0.175$ estimated from
 eq.(\ref{finaleq}). As  the network size increases, $r_{\mbox{av.}}$ at small
 coupling
 decreases, which suggests that the finite $r_{\mbox{av.}}$ at small coupling
 is the finite-size
 effect. The order parameter begin to increases rapidly above $K_c$. We
 note that by increasing the system size,
 the increase of order parameter begins at smaller coupling. This
 means that the critical coupling  $K_c$ decreases as the system
 size increases. We also show $K_c$ estimated from eq.(\ref{finaleq}) in
 this figure. The estimated $K_c$ qualitatively coincide with the
 coupling constant at which the order parameter increases rapidly. We
 conclude that our analysis and the results of the numerical simulation
 show a good  agreement also in the random scale-free network. These results
 suggest that in the infinite size scale-free network the critical
 coupling constant $K_c$ becomes zero, just same as the continuum limit
 equation.

 To compare  the results of  the numerical simulation and the analysis
 more precisely, we need a more accurate estimation of $K_c$ from the 
 numerical simulation.  In the case of the globally coupled networks  and
 Watts-Strogatz model, $K_c$ is numerically obtained as the point at which
 $N^{0.25} r_{\mbox{av.}}$ becomes  independent of  the size of the
 network\cite{Hong}.  In their analysis, there exists an assumption that
 $K_c$ does not depend on the size of the network. On the other hand,
 our analysis and simulation show that $K_c$ depends clearly on the size of
 the network through the average of the square of the degrees. Therefore
 it  is impossible to obtain accurate $K_c$  from  finite-size analysis.
 The exact estimation of $K_c$ is a difficult task.

\begin{figure}[t]
  \resizebox{.4\textwidth}{!}{\includegraphics{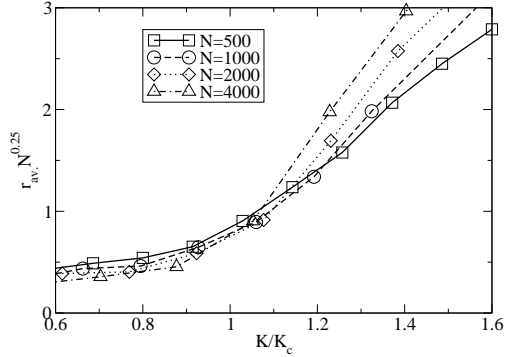}}
  \caption{The relation between $N^{0.25} r_{\mbox{av.}}$ and $K/K_c$, where
  $K_c$ is the value estimated from eq.(\ref{finaleq}).\label{scalefree-finite-size-scaling}}.

 \end{figure}

 However, we find that $K_c$ derived from eq.(\ref{finaleq}) seems to have
 a strong relation to the phase transition. We rescale $K$ 
 by $K_c$ which is obtained from eq.(\ref{finaleq}), and plot the
 relation between $N^{0.25} r_{\mbox{av.}} $ and $K/K_c$ in 
 Fig.\ref{scalefree-finite-size-scaling}. In the case of
  $N=1000, 2000$ and $4000$,
  the well-defined crossing point exists  at $K/K_c=1$. In
 the case of $N=500$, $N^{0.25}r_{\mbox{av.}}$  at $K/K_c=1$ is a little
 larger than in the other cases. However, this difference is small, and it
 seems that $K/K_c=1.0$ is the crossing point at large $N$. This result
 is similar to the results of the finite-size scaling in the
 globally coupled networks and Watts-Strogatz model. In these models,
 there exists a crossing point at $K=K_c$.
 On the other hand, our analysis is not a precise determination of the 
 critical coupling strength. To avoid the size dependence of $K_c$, we 
rescale $K$ and  we have no guarantee that such a  rescaling is
 valid for  scale-free network model. However, our result strongly
 suggests that $K_c$ obtained from the numerical simulation coincides with
 the result of the analytic solution.

 To conclude this section, we carried out the simulations on
 Erd\H{o}s-R\'{e}nyi model and the random scale-free network. All the
 results of these simulations show a qualitative agreement with the
 analysis in the previous section.

\section{Summary and Discussion}

 In this paper, we study the frequency synchronization of the random oscillator
 network.
 By analyzing the  continuum limit  equation, we find  that mean-field type
 synchronization occurs in
 random network model. We obtain the sufficient condition for the
 synchronization. Especially we find  that 
   the threshold for the synchronization is absent in scale-free random
 network if  $2 < \gamma \le 3$.
 The results of numerical simulations are in good agreement with this
 analysis.

 One of the most astonishing result in the dynamics of scale-free  network is
 the absence of epidemic threshold in SIS model.  
 Our result  seems to be similar to the result in SIS model,
 however, there exists a large difference between them.
 In SIS model the absence of epidemic threshold is the result of the
 divergence of
 $k_{\mbox{nn}}$, the mean degree of the nearest neighbor
 nodes\cite{SIS}. On the other
 hand,
 in our model the absence of threshold originates from the degree-dependence
 of 
 the coupling between order parameter and oscillators.
   The coupling between the oscillators and the mean-field
 is proportional to the degree of the nodes, as shown in eq.(\ref{mf-eq}), and 
the contribution to the order parameter from the oscillator is also
 proportional to the degree of the node, shown in eq.(\ref{mf}).
 These degree-dependence result in the $k^{2}$-dependence of
 eq.(\ref{critical}), which leads to the absence of the threshold in random
 scale-free network.
 Therefore
 there is a large difference between the absence of threshold in SIS model
 and the synchronization, though these are apparently
 similar results.
 To clarify this difference, we will need to study the
 synchronization in the other network models.  As Egu\'{\i}luz
 and Klemm has shown, scale-free network with large clustering
 coefficient has
 epidemic threshold in SIS model\cite{Eguiluz}, due to the smallness of
 $k_{\mbox{nn}}$. The different behavior of the threshold may appear in our 
 oscillator network, because the absence of threshold is not caused by the
 divergence of $k_{\mbox{nn}}$, but by degree dependence of
 mean field-oscillator  coupling. The study of the synchronization in
 other scale-free network models is a future problem.

\begin{acknowledgements}
 We acknowledge to Y. Nishiura, M. Iima and T. Yanagita for fruitful comments. 
\end{acknowledgements}

\end{document}